\documentclass{elsart}
\usepackage{psfig}
\usepackage{natbib}

\def\asca{{\it ASCA\/}}

\def\rosat{{\it ROSAT\/}}
\def\sax{{\it BeppoSAX}}
\def\xmm{{\it XMM\/}}

\def\simgt{\lower 2pt \hbox{$\, \buildrel {\scriptstyle >}\over {\scriptstyle
\sim}\,$}}
\def\simlt{\lower 2pt \hbox{$\, \buildrel {\scriptstyle <}\over {\scriptstyle
\sim}\,$}}

\begin{document}

\runauthor{Brandt}


\begin{frontmatter}

\title{Observational Similarities and Potential Connections Between 
Luminous Ultrasoft NLS1s and BALQSOs}

\author[PSU]{W.N. Brandt}
\author[PSU]{S.C. Gallagher}
\address[PSU]{Department of Astronomy \& Astrophysics, 525 Davey Lab, 
The Pennsylvania State University, University Park, Pennsylvania 16802 USA}

\begin{abstract}
Luminous ultrasoft NLS1s and low-ionization BALQSOs share many properties,
and they both represent important extremes of the active galaxy phenomenon. 
We briefly discuss their observational similarities as well as potential 
physical connections between them, concentrating on the X-ray point of view. 
We
present several ways by which potential connections might be further tested. 
\end{abstract}


\end{frontmatter}


\section{Introduction}

It is clear from the papers presented at this workshop that studies
of ultrasoft Narrow-Line Seyfert~1 galaxies (NLS1s) have undergone
exciting growth over the past few years. Progress has been made
defining their phenomenological properties; NLS1s show extreme
spectral shapes and variability at a variety of wavelengths 
(e.g., X-ray, optical and radio). While a solid physical understanding 
of the origin of their extreme properties has not yet emerged, they 
are plausibly objects with high values of the mass
accretion rate relative to the Eddington rate ($\dot M/\dot M_{\rm Edd}$). 

In this paper, we will briefly discuss potential connections between 
luminous NLS1s and Broad Absorption Line QSOs (BALQSOs), concentrating on 
the X-ray point of view. We will informally 
advocate such connections to the greatest extent possible in an attempt to 
stimulate further work in this area. In the X-ray band, even the basic 
phenomenological
properties of BALQSOs are poorly known (see \S2). Because of their low X-ray 
fluxes, the present observations do not allow one to draw a characteristic 
X-ray spectral energy distribution for a BALQSO. However, with the advent of 
the new generation of X-ray observatories, one hopes that X-ray studies 
of BALQSOs will undergo exciting growth in the coming decade
comparable to that achieved for NLS1s in the past few years. 


\section{X-rays from Broad Absorption Line QSOs}

Most luminous QSOs are thought to have BAL outflows which 
comprise a major part of the nuclear 
environment (e.g., Weymann 1997, hereafter W97). 
Furthermore, the BAL phenomenon has interesting connections with the
`radio volume control' of QSOs (e.g., \S4 of Becker et~al. 2000 and
references therein), and BAL outflows may clear gas from
QSO host galaxies and thereby affect star formation and QSO fueling
over cosmic timescales (e.g., Fabian 1999). With X-rays, one would
like to study the BAL wind in absorption to determine properties
such as its column density, ionization state, abundances, and covering 
factor. One would also like to probe the nuclear geometry (e.g., using 
the iron~K$\alpha$ line and X-ray variability) and determine if 
BALQSOs have normal QSO X-ray continua underlying their absorption. 

\begin{figure}[htb]
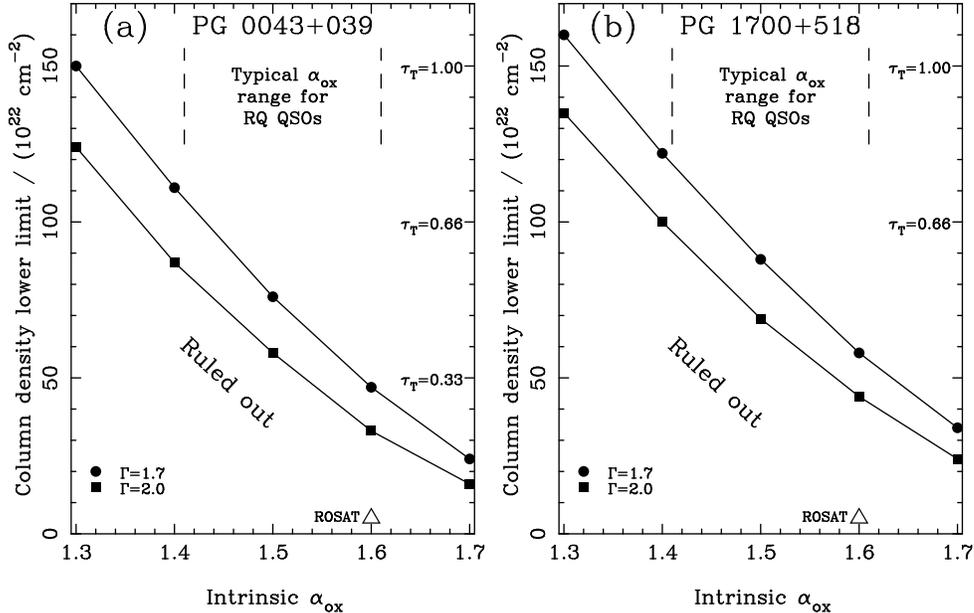

\centerline{
\psfig{figure=brandt_fig_1a.ps,height=3.2truein,width=2.5truein,angle=0}
\psfig{figure=brandt_fig_1b.ps,height=3.2truein,width=2.5truein,angle=0}}
\caption{
Column density lower limits for (a) PG~0043+039 and 
(b) PG~1700+518 derived using \asca\ SIS0 data. We 
show the inferred column density lower limit as 
a function of the intrinsic (i.e., absorption-corrected) value of
$\alpha_{\rm ox}$ (the slope of a nominal power law connecting the rest-frame
flux density at 2500~\AA\ to that at 2~keV). The square data points
are for an X-ray photon index of $\Gamma=2.0$, and the circular
dots are for an X-ray photon index of $\Gamma=1.7$. The open triangle
at $\alpha_{\rm ox}=1.6$ illustrates the typical BALQSO column 
density lower limit found based on \rosat\ data. The numbers
along the right-hand sides of the panels show the Thomson optical depth
of the corresponding column density. The column density lower limits 
shown in this plot are for absorption by neutral gas with solar
abundances. Ionization of the gas can significantly raise the 
required column density. 
PG~0043+039 has many characteristics typical of low-ionization 
BALQSOs, but at present there is no convincing evidence for BALs 
due to low-ionization transitions 
(see Turnshek et~al. 1994 for details). 
PG~1700+518 is a low-ionization BALQSO. 
From Gallagher et~al. (1999).}
\end{figure}

Studies with \rosat, \asca, and \sax\ have established that
BALQSOs are very weak emitters in the 0.1--2.0~keV band, and that
with a few notable exceptions they are also weak when studied
with more penetrating 2--10~keV X-rays (e.g., Green \& Mathur 1996;
Gallagher et~al. 1999; Mathur et~al. 2000; see Brandt et~al. 2000
for a review). The higher detection fraction obtained in hard 
X-rays ($\approx 50$\%) is generally consistent with the presence
of heavy internal X-ray absorption. If BALQSOs have normal 
underlying QSO X-ray continua, then large column densities of 
$\simgt 5\times 10^{23}$~cm$^{-2}$ are required to extinguish the 
X-ray emission in several cases (see Figure~1). 
These large X-ray column densities have important 
physical implications. For example, if the X-ray
absorption occurs in gas at a distance $\simgt 3\times 10^{16}$~cm 
outflowing with a significant fraction of the terminal velocity 
measured from the UV BALs, large mass outflow rates 
($\dot M_{\rm outflow}\simgt 5$~M$_\odot$~yr$^{-1}$;
$\dot M_{\rm outflow}\simgt \dot M_{\rm accrete}$) and kinetic 
luminosities ($L_{\rm kinetic}\simgt L_{\rm ionizing}$) are derived. 
While these can be reduced if the X-ray and UV absorbers in 
BALQSOs differ, the possibility of such powerful outflows 
demands further study. 

\begin{figure}[htb]
\centerline{\psfig{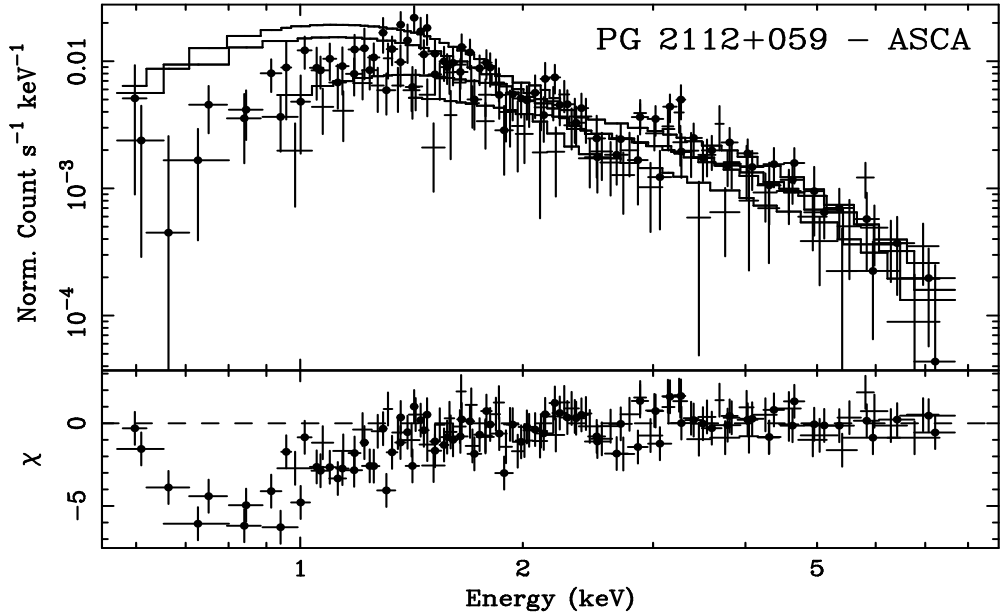}}
\caption{ASCA SIS and GIS observed-frame spectra of the BALQSO 
PG~2112+059. A power-law model has been fit to the data above
2~keV and then extrapolated back to show the deviations at
low energies. The ordinate for the lower panel (labeled $\chi$)
shows the fit residuals in terms of $\sigma$ with error bars
of size one. Note the systematic absorption residuals at low
energies. From Gallagher et al. (2000, in preparation).}
\end{figure}

At present, it has been possible to measure the intrinsic X-ray 
continuum shape well for only one `bona-fide' BALQSO, PG~2112+059
($z=0.46$; Gallagher et~al. 2000, in preparation; see 
Jannuzi et~al. 1998 for the UV spectrum). This 
object is one of the most luminous
PG~QSOs at $z<0.5$ with $M_{\rm V}=-27.3$. Our recent ASCA 
spectra show strong absorption below $\approx 2$~keV, but the
rest-frame 3--10~keV continuum appears relatively unaffected by 
absorption (see Figure~2). We measure a power-law photon index of 
$\Gamma_{3-10}=1.94^{+0.23}_{-0.21}$ (90\% confidence). This 
first detection of a typical power-law X-ray continuum for a 
BALQSO is reassuring and argues that BALQSOs indeed have normal 
X-ray power sources underneath their absorption. 


\section{Observational Comparison of NLS1s and BALQSOs}

Several papers have qualitatively commented on observational
similarities between NLS1s\footnote{The main interest here is
in the majority of NLS1s that lie toward the strong Fe~{\sc ii}, 
weak [O~{\sc iii}] end of Boroson \& Green (1992, hereafter BG92) 
eigenvector~1 (i.e., the `I~Zwicky~1 objects'). We use the term `NLS1s' 
loosely to refer to such objects.} and BALQSOs 
(e.g., Boroson \& Meyers 1992, hereafter BM92; Brandt et~al. 1997;
Laor et~al. 1997; Lawrence et~al. 1997; Leighly et~al. 1997). 
Two demographic facts must immediately be considered:
(1) NLS1-type objects are found over a wide range of luminosity ranging 
from low-luminosity Seyferts such as NGC~4051 to luminous QSOs such as 
PHL~1092, while BALs are seen only in luminous QSOs, and
(2) most and perhaps all QSOs have BAL regions while most QSOs
do not show the characteristic `NLS1' properties. 
The first suggests that only the luminous NLS1s may be 
unified with BALQSOs, and the rest of this paper shall 
focus on such objects. 
Fact~2 is double-edged: while it is implausible to try to unify 
{\it all\/} BALQSOs and luminous NLS1s, it appears that there
must be {\it some\/} connection between them. The objective 
is then to identify the subset of BALQSOs that are the `cousins' 
of luminous NLS1s. 

Based on a comparison of Na~{\sc i} emission in I~Zwicky~1 and
BALQSOs, BM92 suggested that ``I~Zwicky~1 is
a promising candidate for a Mg~{\sc ii} BALQSO in which our line
of sight does not pass through a BAL cloud.'' Low-ionization 
BALQSOs, showing absorption by species such as Mg~{\sc ii} and 
Al~{\sc iii}, comprise 10--15\% of the BALQSO population and
often have reddened continua and extremely weak 
[O~{\sc iii}] emission (e.g., Weymann et~al. 1991, hereafter W91; 
BM92; Sprayberry \& Foltz 1992). They are thought to have large 
amounts of relatively cool gas and dust in their nuclei, and 
they are sometimes postulated to be young, recently activated QSOs. 
It appears that low-ionization BALQSOs have larger BAL-region covering 
factors ($f_{\rm c, BAL}$) than typical 
BALQSOs (e.g., BM92; Turnshek et~al. 1997). 

Luminous NLS1s and low-ionization BALQSOs appear to 
share many common properties, lending plausibility to a connection
between them. These are listed in Table~1, where a few relevant 
references for the low-ionization BALQSO properties are also given. 
In several cases the low-ionization BALQSO properties have only 
been measured in small and statistically incomplete samples;
better systematic studies are clearly needed so that one can
reason armed with more than anecdotal evidence. Systematic
H$\beta$ FWHM measurements are in particular required. The four
low-ionization BALQSOs in BM92 have H$\beta$ FWHM of
$\approx 2000$--3000~km~s$^{-1}$, fairly narrow for high 
luminosity objects. However, there also appear to be at
least several low-ionization BALQSOs that have substantially
broader H$\beta$ FWHM of 4000--6000~km~s$^{-1}$ 
(e.g., McIntosh et~al. 1999; M. Brotherton, private communication). 
While a few radio-loud NLS1s and low-ionization BALQSOs exist
(e.g., Becker et~al. 2000; Grupe et~al. 2000), both classes 
are in general conspicuously radio quiet relative to Broad-Line 
Seyfert~1s and high-ionization BALQSOs. 

\begin{table*}
\caption{Selected Multiwavelength Properties of NLS1s and Low-Ionzation 
BALQSOs}
\label{defparagcl} 
\begin{center}
\begin{tabular}{l l l}
Property                        & NLS1s                 & Low-Ionization 
BALQSOs \\
\hline 
[O~{\sc iii}] luminosity        & Low                   & Low (BM92; Turnshek 
et~al. 1997) \\
Optical Fe~{\sc ii}/H$\beta$    & Large                 & Large$^\star$ (BM92; 
Lawrence et~al. 1997) \\
Balmer line asymmetry           & Strong blue wings     & Strong blue wings 
(BM92) \\
H$\beta$ FWHM                   & Small                 & See the text of \S3 
(BM92) \\
Radio loudness                  & Generally quiet       & Generally quiet 
(W97; Becker et~al. 2000) \\
Far-infrared luminosity         & Generally high        & Generally high (Low 
et~al. 1989; W91) \\
\hline 
\end{tabular}
\end{center}
\vspace*{.6cm}
\noindent
$^{\star}$ The low-ionization BALQSOs also have stronger UV Fe~{\sc ii} 
emission than non-BALQSOs (W91). \\
\end{table*}

\begin{figure}[htb]
\centerline{
\psfig{figure=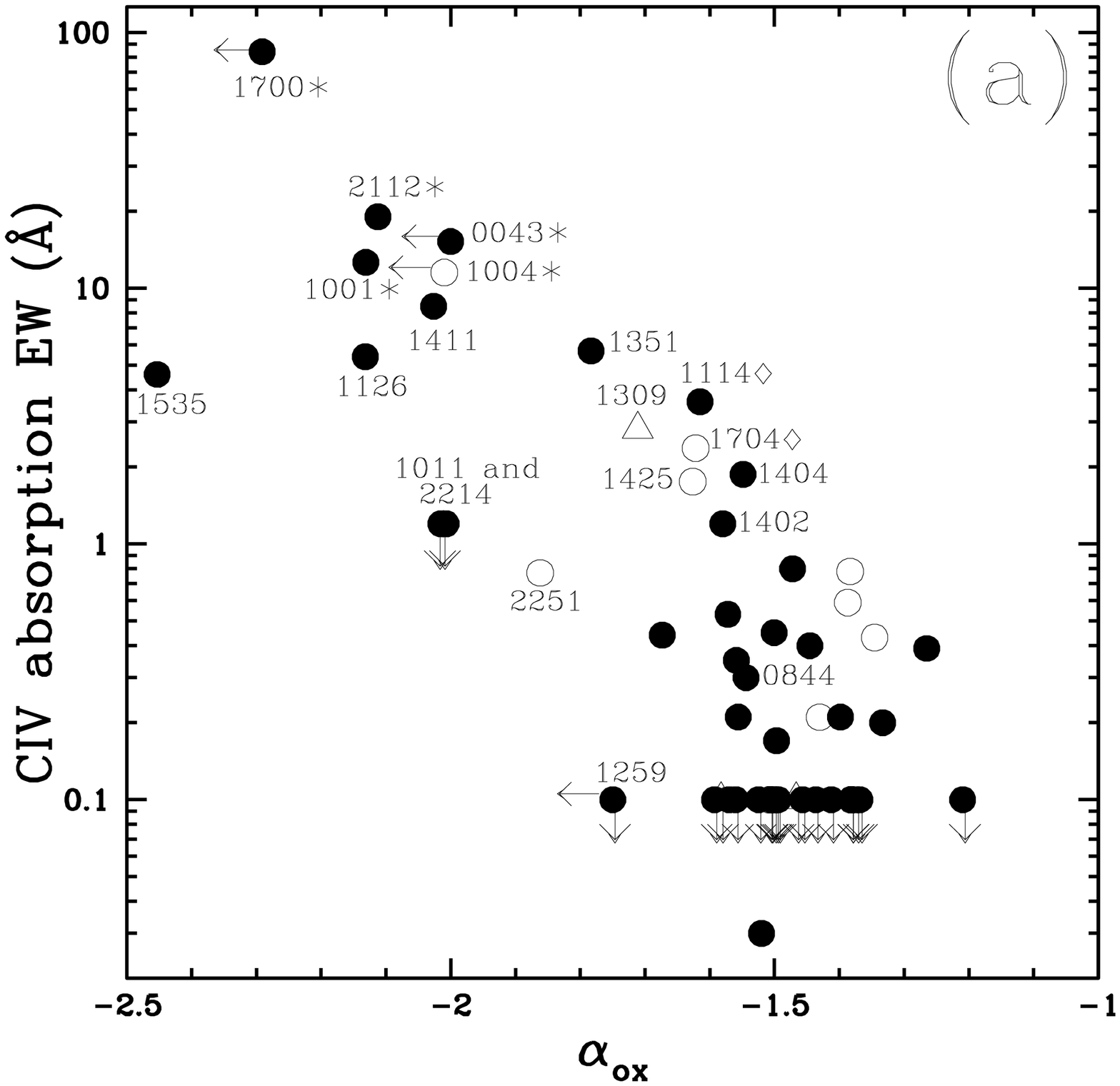,height=3.0truein,width=2.8truein,angle=0}
\psfig{figure=brandt_fig_3b.ps,height=2.8truein,width=2.85truein,angle=-90}}
\caption{
(a) Plot of $\alpha_{\rm ox}$ versus C~{\sc iv} $\lambda 1549$ absorption-line
EW (in the rest frame) for the BG92 QSOs with UV coverage. Following BG92, 
solid dots are radio-quiet QSOs, open triangles are core-dominated radio-loud
QSOs, and open circles are lobe-dominated radio-loud QSOs. Particularly
interesting QSOs are labeled by the right ascension part of their name;
note the extreme location of PG~1535+547. An asterisk to the right of an 
object's name indicates that it is a BALQSO or probable BALQSO, and a
diamond indicates that it is known to have an X-ray warm absorber.
>From Brandt, Laor \& Wills (2000).  
(b) ASCA SIS and GIS observed-frame spectra of PG~1535+547. A power-law
model with absorption partially covering the X-ray continuum is 
shown and provides an acceptable fit to the data. The ordinate 
for the lower panel is as per Figure~2. From 
Gallagher et~al. (2000, in preparation).}
\end{figure}

Empirically, luminous NLS1s and low-ionization BALQSOs appear quite 
different in X-rays. Luminous NLS1s are strong emitters of soft X-rays 
while low-ionization BALQSOs are very weak in this band (see \S2). 
However, this may just be due to absorption in low-ionization BALQSOs 
that prevents us from seeing the underlying X-ray continuum; the
intrinsic X-ray continua of low-ionization BALQSOs could be 
very similar to those of NLS1s. If better data eventually
show, for example, that many low-ionization BALQSOs have unusually
steep X-ray power laws with photon indices of $\Gamma=2.0$--2.5, this
would strongly suggest a connection with luminous NLS1s (compare with
Brandt, Mathur \& Elvis 1997).\footnote{PG~2112+059, the only BALQSO
with a measured X-ray continuum shape (see \S2), does not 
have published spectral coverage of the key Mg~{\sc ii} region
for determining if it is a high-ionization or low-ionization BALQSO.} 

It appears likely that absorbing gas capable of completely
obscuring the intrinsic X-ray continuum is present in the nuclei of 
some NLS1s. A range of absorption strengths is already known to 
be present and is being extended to nearly BALQSO-level column 
densities with new observations. Several NLS1s are 
known to have X-ray warm absorbers (e.g., Brandt et~al. 1997), and
`anomalous' absorption around 1~keV may be present in some NLS1s as
well (e.g., Leighly et~al. 1997). The 1~keV absorption features have
been speculated to be associated with highly ionized gas
($N_{\rm H}\sim 2\times 10^{21}$~cm$^{-2}$) flowing out of the 
nucleus at 0.2--0.3$c$. However, other explanations are also plausible 
(e.g., Nicastro, Fiore \& Matt 1999), and the evidence requires 
substantial improvement. Heavier absorption still is known to be present 
in Mrk~507 (Iwasawa, Brandt \& Fabian 1998) and the soft X-ray weak
PG~1535+547 (Mrk~486; Gallagher et~al. 2000, in preparation; also see
Brandt, Laor \& Wills 2000). The X-ray absorption in Mrk~507 is
complex but appears to be mainly due to mildly ionized gas with a 
column density of $\approx$~(2--3)$\times 10^{21}$~cm$^{-2}$. 
Better UV spectra are now needed to search for UV absorption
lines from this object. PG~1535+547 shows the 
heaviest X-ray absorption known in a NLS1; modeling
it with neutral gas requires a column density of 
$\approx 1.2\times 10^{23}$~cm$^{-2}$ and a covering fraction 
of $\approx 90$\% (see Figure~3). Unfortunately, the photon
index of the underlying power law is poorly constrained at
present ($\Gamma=2.02^{+0.92}_{-0.95}$) due to limited photon
statistics; spectra from \xmm\ should reveal if the power law is
unusually steep. The characteristics of the X-ray absorption 
are likely related to the observed UV absorption and high polarization 
(e.g., Smith et~al. 1997), and the strength of the X-ray absorption
is remarkable in a type~1 object. The derived column density is close 
to that seen for BALQSOs as well as the tori of some Seyfert~2 galaxies. 
In another example, based on a steep X-ray power-law continuum and 
strong variability, Iwasawa et~al. (1996) have suggested that an 
obscured NLS1 nucleus may reside in the Seyfert~2 galaxy IRAS~18325--5926.


\section{Physical Interpretation}

X-ray studies of ultrasoft NLS1s suggest that they are characterized
by extreme values of a primary physical parameter, probably
$\dot M/\dot M_{\rm Edd}$ (e.g., Brandt \& Boller 1999). Observations
at optical and UV wavelengths have also revealed that NLS1s
have large amounts of dense ($\sim 10^{11}$~cm$^{-3}$), low-ionization,
line-emitting gas in their Broad Line Regions (BLRs; e.g., BG92; 
Wills et~al. 1999; Kuraszkiewicz et~al. 2000). However, the physical
relation between high $\dot M/\dot M_{\rm Edd}$ and dense BLRs remains
unclear. Wills et~al. (1999) suggest that gas injected into the nucleus 
by starburst activity may lead to both high $\dot M/\dot M_{\rm Edd}$ 
and high BLR gas densities. 

The most straightforward way to connect luminous NLS1s and low-ionization 
BALQSOs would be to postulate that both are characterized by high 
$\dot M/\dot M_{\rm Edd}$. It would then just be the presence or absence 
of BAL material along the line of sight that determines the classification 
of a given high $\dot M/\dot M_{\rm Edd}$ object. The amount of 
line-of-sight material could naturally be set by the nuclear 
orientation, although random fluctuations in the number of intervening 
BAL `clouds' could perhaps also be relevant. The average value of 
$f_{\rm c, BAL}$ for high $\dot M/\dot M_{\rm Edd}$ objects would
set the relative number densities of luminous NLS1s and low-ionization 
BALQSOs. While $\langle f_{\rm c, BAL} \rangle$ for high
$\dot M/\dot M_{\rm Edd}$ objects could plausibly be as large as 
$\sim 50$\% given likely selection effects against low-ionization
BALQSOs (e.g., Goodrich 1997; Krolik \& Voit 1998), a value much 
larger than this would probably overpredict the
number of low-ionization BALQSOs relative to luminous NLS1s. 

The scenario of the previous paragraph has a couple of attractive features. 
First of all, the strong, high $f_{\rm c, BAL}$ outflows of low-ionization 
BALQSOs might well result from high $\dot M/\dot M_{\rm Edd}$. Accreting 
systems with high $\dot M/\dot M_{\rm Edd}$ have a greater ability to 
radiatively drive outflows due to their larger photon luminosities per unit 
gravitational mass. Furthermore, radiation-dominated analogues of advective 
inflow-outflow solutions may become important at high $\dot M/\dot M_{\rm 
Edd}$
(e.g., \S5 of Blandford \& Begelman 1999; Blandford \& Begelman, in 
preparation). 
The scenario of the previous paragraph may also be able to
explain why NLS1s have unusually dense BLRs without requiring 
a secondary agent, such as starburst activity. Some of the gas 
accelerated as a result of the high value of $\dot M/\dot M_{\rm Edd}$ 
is likely to end up in the BLR, where it can cool and increase the
local density (see \S8.2 of Brandt, Laor \& Wills 2000). 

If high-velocity outflows with large covering factors are present 
in the nuclei of all luminous active galaxies with high 
$\dot M/\dot M_{\rm Edd}$, one might hope to occasionally see
`occultations' when material moves through the
line of sight. Such occultations have been seen in several
high $\dot M/\dot M_{\rm Edd}$ X-ray binaries where they can cause 
dramatic X-ray variability, sometimes without strong spectral 
changes (e.g., Dower, Bradt \& Morgan 1982; Brandt et~al. 1996). 
An intriguing, speculative possibility is that {\it some\/} of the 
extreme X-ray variability seen in the most variable NLS1s is due to 
occultations. The large-amplitude variability seen in IRAS~13224--3809, 
for example, is difficult to explain as true luminosity changes on 
energetic grounds. An occultation model is able to explain the 
variability seen on a timescale of days provided the occulting 
material is quite thick and moves rapidly across the line of sight
(Boller et~al. 1997). While BAL winds are both thick and rapidly
moving, they probably could not cause the variability because
they originate too far from the nucleus. However, highly ionized 
material launched at smaller radii (e.g., the `hitchhiking gas' 
of Murray et~al. 1995) may be relevant. 


\section{Future Prospects for X-ray Studies}

Observations with the new generation of X-ray observatories should
be able to test some of the ideas discussed above. For example, 
systematic measurements of the X-ray continuum shapes of both 
high-ionization and low-ionization BALQSOs can test if low-ionization 
BALQSOs have unusually steep power-law continua with photon indices 
of $\Gamma=2.0$--2.5. If they do, this would immediately suggest a 
connection with luminous NLS1s and high $\dot M/\dot M_{\rm Edd}$, 
and it is encouraging to see that some progress is now being made 
measuring BALQSO X-ray continua (see \S2 and \S3). Unusually strong 
X-ray variability from low-ionization BALQSOs would also suggest a 
connection with luminous NLS1s. 
Furthermore, high-quality X-ray and UV spectroscopy of moderately 
absorbed NLS1s should help to bridge the gap between unabsorbed NLS1s 
and the heavily absorbed BALQSOs. For at least one NLS1, PG~1535+547, 
the X-ray column density is comparable to that inferred for
BALQSOs, and UV absorption is also present (see \S3). 
The NLS1s with the strongest known absorption (e.g., Mrk~507 and
PG~1535+547) unfortunately fall short of having QSO-level luminosities;
hopefully more luminous examples will be found. 

Finally, if low-ionization BALQSOs and luminous NLS1s are indeed related, 
one might hope to see `transitions' between these two types of 
objects. These could occur due to abrupt changes in the 
flow geometry of the absorbing material. Strong changes in the 
absorption-line properties of some low-ionization BALQSOs have already been 
seen (e.g., Boroson et~al. 1991; Junkkarinen, Cohen \& Hamann 1999), although 
these appear to occur on fairly long timescales. Perhaps at the Y3K workshop
on NLS1s there will be a celebration when Mrk~231 undergoes such a
transition and ends up looking something like PHL~1092. This would be
a sight to rival the Y2K launch of \xmm! 


\section{Acknowledgments}

We acknowledge the support of NASA LTSA grant NAG5-8107 (WNB)
and the Pennsylvania Space Grant Consortium (SCG). We thank B.J. Wills
for helpful comments.



\end{document}